\documentclass[aps,prb,twocolumn,showpacs,superscriptaddress,am,floatfix]{revtex4-1}
\usepackage{graphicx}
\usepackage{amsmath}
\usepackage{amssymb}
\usepackage{color}

\newcommand{\bainiro}{Ba$_3$InIr$_2$O$_9$}

\newcommand{\abmo}{A$_3$B'M$_2$O$_9$}

\begin{document}
\title{Soft and anisotropic local moments in 4$d$ and 5$d$ mixed-valence M$_2$O$_9$ dimers}

\author{Ying Li$^*$} 
\affiliation{Department of Applied Physics and MOE Key Laboratory for Nonequilibrium Synthesis and Modulation of Condensed Matter, School of Science, Xi'an Jiaotong University, Xi'an 710049, China}
\affiliation{Institut f\"ur Theoretische Physik, Goethe-Universit\"at Frankfurt,
Max-von-Laue-Strasse 1, 60438 Frankfurt am Main, Germany}
\author{Alexander A. Tsirlin}
\author{Tusharkanti Dey$^{\dagger}$}
\author{Philipp~Gegenwart}
\affiliation{Experimental Physics VI, Center for Electronic Correlations and Magnetism, University of Augsburg, 86159 Augsburg, Germany}
\author{Roser Valent{\'\i}$^{\ddagger}$}
\affiliation{Institut f\"ur Theoretische Physik, Goethe-Universit\"at Frankfurt,
Max-von-Laue-Strasse 1, 60438 Frankfurt am Main, Germany}
\author{Stephen M. Winter$^{\mathsection}$}
\affiliation{Institut f\"ur Theoretische Physik, Goethe-Universit\"at Frankfurt,
Max-von-Laue-Strasse 1, 60438 Frankfurt am Main, Germany}
\date{\today}

\begin{abstract}

        We investigate via exact diagonalization
	of finite clusters the electronic
	structure and magnetism of M$_2$O$_9$ dimers in the mixed-valence
	hexagonal perovskites {\abmo} for various different fillings of 
	4$d$ and 5$d$ transition-metal M ions. 
	We find that the magnetic moments of such dimers are 
	determined by a subtle interplay of spin-orbit coupling, Hund's
	coupling, and Coulomb repulsion, as well as 
	the electron filling of the M ions.
	Most importantly, 
	the magnetic moments are anisotropic and 
	temperature-dependent.  This behavior is a result
	of spin-orbit coupling, magnetic
	field effects, and
	the existence of several nearly-degenerate electronic
	configurations whose proximity allows occupation of excited states
	already at room temperature. This analysis is consistent with 
	experimental susceptibility measurements for a variety of dimer-based
	materials. Furthermore,
	we perform a survey of {\abmo} materials and propose ground-state
	phase diagrams for the experimentally relevant M fillings of $d^{4.5}$, 
	$d^{3.5}$ and $d^{2.5}$. Finally,
	our results show that the
	usually applied Curie-Weiss law with a constant magnetic moment cannot
	be used in these spin-orbit-coupled materials. 

\end{abstract}
\maketitle
\par

\section{Introduction}

Oxides of $4d$ and $5d$ transition metals feature strong spin-orbit coupling,
 moderate electronic correlations and sizable
metal-oxygen hybridization. Together, these effects may give rise to exotic
spin-orbital states, including quantum spin liquids relying on strongly
anisotropic effective magnetic Hamiltonians~\cite{Witczak-Krempa2014,Rau2016,Schaffer2016,Winter2017,Cao2018}. An interesting class of such
materials is the 6H perovskite family {\abmo} (M = Ir, Os, Re, Rh,
Ru), which features face-sharing M$_{2}$O$_9$ dimers as the central magnetic and structural
unit. Whereas the A-cations are almost always divalent, the oxidation state of
the B'-cations ranges from 1+ to 4+, thus imposing different charge states to
the transition metal M. Compounds with integer oxidation states of M usually
behave as  common
spin dimers~\cite{darriet1976,chen2019}, where two magnetic centers
are coupled by a moderately strong exchange interaction leading to a
non-magnetic ground state. However, in $5d^4$ systems with the non-magnetic
($J_{\rm eff}=0$) state of individual M ions, unusual effects like
``excitonic'' magnetism can be expected~\cite{kim2017} if interdimer magnetic couplings
exceed the energy gap to excited multiplets with $J_{\rm eff} > 0$. The
Ba$_3$B'Ir$_2$O$_9$ iridates with divalent B' = Zn, Ca, Sr serve as possible
experimental examples for this scenario~\cite{nag2016origin,nag2019hopping}. 

In contrast, if the total charge of [A$_3$B'] is odd, it induces a half-integer
oxidation state in M, i.e., a mixed (or intermediate) valence of the
transition-metal ion~\cite{Byrne1970}. In this case, having an odd number of electrons per dimer
ensures a finite spin moment in the ground state. Various different valencies
are possible including $d^{4.5}$ in iridates with trivalent B'~\cite{Doi2004,Sakamoto2006}, $d^{3.5}$ in ruthenates with trivalent B'~\cite{DOI2002,Senn2013,Ziat2017} or iridates with monovalent B'~\cite{KIM2004}, and $d^{2.5}$ in ruthenates and osmates with monovalent B'~\cite{Stitzer2002,Stitzer2003}. Such materials
have been less explored theoretically, despite the fact that their
experimental magnetic response reveals several peculiarities. At high
temperatures, where the mixed-valence dimers can be seen as isolated, magnetic
susceptibility deviates from the conventional Curie-Weiss
behavior~\cite{DOI2002,Sakamoto2006,SHLYK2007} suggesting a non-trivial
temperature evolution of the local magnetic moment. At low temperatures,
interactions between the dimers become important, and signatures of frustrated
magnetic behavior~\cite{Ziat2017} including possible formation of a
spin-liquid ground state~\cite{Dey2017} have been reported.  

As we discuss in this work, even on the level of a single dimer~\cite{Streltsov2016, Streltsov2014, Kugel2015, Khomskii2016} a variety of different local states can be realized as a function of electronic filling, dimer geometry, and spin-orbit coupling strength. Understanding the magnetic models describing interactions between such dimers first requires an understanding of the local electronic structures of an individual dimer. Here, we endeavor to obtain a microscopic insight into the electronic state and magnetism
of the mixed-valence M$_2$O$_9$ dimers with different fillings $d^{4.5}$, $d^{3.5}$ and $d^{2.5}$ encountered in $4d/5d$ hexagonal perovskites. In particular, we focus on experimentally relevant
details, such as the relation between ground state and paramagnetic
susceptibility. Performing exact diagonalization (ED) of dimer clusters, we find
a sizable state-dependent uniaxial anisotropy of the magnetic moment and its
temperature dependence, which can be employed to identify different ground
states from experiment. 

This paper is organized as follows. In Section \ref{sec:details}, we first
provide a preliminary description of the local Hamiltonian for each dimer, followed
by the definition of the effective magnetic moment in Sec.~\ref{sec:moments}. In Sections
\ref{sec:d45}--\ref{sec:d25}, we consider the ground states and behavior of
the effective moment for dimers with $d^{4.5}$, $d^{3.5}$, and $d^{2.5}$
filling and in Section \ref{conclusions} we present our conclusions.

%
%
%
%

\section{Dimer Model}\label{sec:details}
\subsection{Electronic Hamiltonian per Dimer}
To first approximation, the hexagonal perovskites, {\abmo} crystallize in the space group $P6_3/mmc$, in which each dimer has local $D_{3h}$ symmetry. The zero-field Hamiltonian for each dimer is given by:
\begin{eqnarray}
\mathcal{H}_{\rm tot} = \mathcal{H}_{\rm hop}+\mathcal{H}_{\rm CF}+\mathcal{H}_{\rm SO} + \mathcal{H}_U
\label{hamil}
\end{eqnarray}
which is the sum of, respectively, the intersite hopping, on-site crystal field, spin-orbit coupling, and Coulomb interactions. We consider only the $t_{2g}$ orbitals $(d_{xy,}, d_{xz}, d_{yz}$) on each metal atom. In this case, the latter term is given by:
\begin{align}
\mathcal{H}_{\rm U} = & U \sum_{i,a} n_{i,a,\uparrow}n_{i,a,\downarrow} + (U^\prime - J_H)\sum_{i,a< b, \sigma}n_{i,a,\sigma}n_{i,b,\sigma} \nonumber \\
 &+ U^\prime\sum_{i,a\neq b}n_{i,a,\uparrow}n_{i,b,\downarrow} - J_H \sum_{i,a\neq b} c_{i,a\uparrow}^\dagger c_{i,a\downarrow} c_{i,b\downarrow}^\dagger c_{i,b\uparrow}\nonumber \\ & + J_H \sum_{i,a\neq b}c_{i,a\uparrow}^\dagger c_{i,a\downarrow}^\dagger c_{i,b\downarrow}c_{i,b\uparrow} 
\end{align}
where $U$ is the on-site Coulomb repulsion, and $J_H$ is the Hund's coupling. The labels $i,j \in \{1,2\}$ refer to metal sites within the dimer, and $a,b \in\{xy,xz,yz\}$ refer to orbitals. Throughout, we use the approximation $U^\prime = U - 2J_H$. 

\begin{figure}[t]
\includegraphics[angle=0,width=0.45\textwidth]{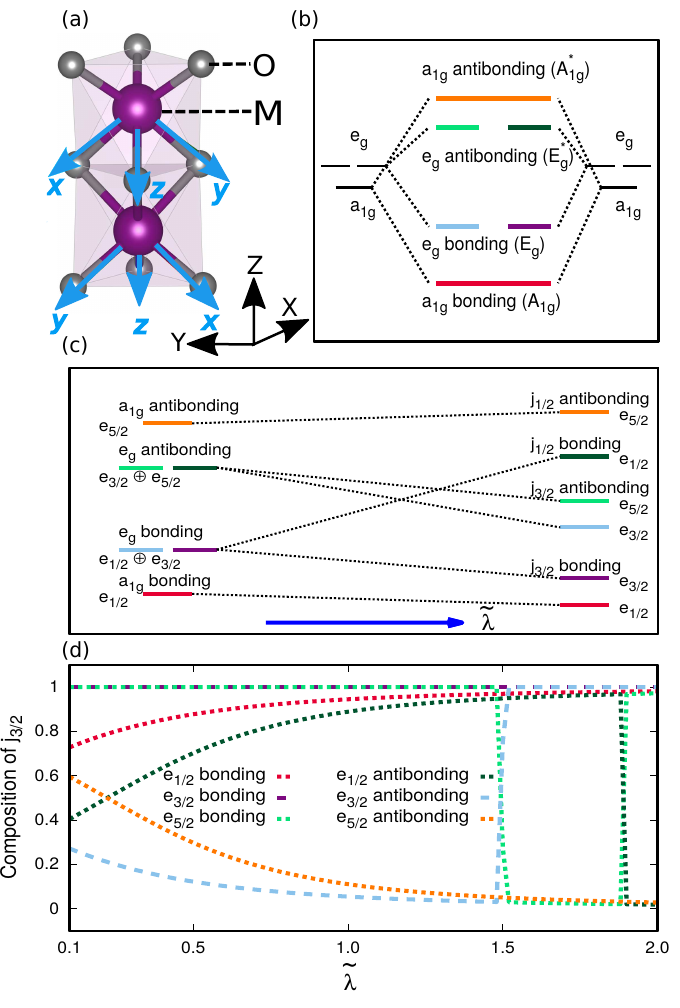} 
\caption{(a) M$_2$O$_9$ face-shared bioctahedral dimer. Shown are
	the local and global coordinates; (b) Energy levels of
	the dimer from ED in the absence of spin-orbit coupling; (c) Energy
	levels-change as a function
	of spin-orbit coupling strength $\tilde{\lambda}=\lambda/t_2$;
	and (d) Composition of $j_{3/2}$ state in terms of
	single-particle levels
	as a function of $\tilde{\lambda}$.} \label{fig:level}
\end{figure}

The single-particle contributions can be written in terms of the local electron creation operators $\mathbf{c}_i^\dagger \equiv ( c_{i,xy,\uparrow}^\dagger \  c_{i,xz,\uparrow}^\dagger \ c_{i,yz,\uparrow}^\dagger\  c_{i,xy,\downarrow}^\dagger \ c_{i,xz,\downarrow}^\dagger \  c_{i,yz,\downarrow}^\dagger) $, with the $t_{2g}$ orbitals defined according to the local coordinates in Fig.~\ref{fig:level}(a).  The local coordinates are defined
 in the basis of the global coordinates as 
\begin{align}
R_1 = 
\left( 
\begin{array}{r@{\hspace{1em}}r@{\hspace{1em}}r}
-\sqrt{6}/6 & -\sqrt{2}/2 & -\sqrt{3}/3 \\
-\sqrt{6}/6 & \sqrt{2}/2 & -\sqrt{3}/3 \\
\sqrt{6}/3 & 0 & -\sqrt{3}/3 \\
\end{array}
\right)
\end{align}
for the first atom and
\begin{align}
R_2 = 
\left( 
\begin{array}{r@{\hspace{1em}}r@{\hspace{1em}}r}
\sqrt{6}/6 & \sqrt{2}/2 & -\sqrt{3}/3 \\
\sqrt{6}/6 & -\sqrt{2}/2 & -\sqrt{3}/3 \\
-\sqrt{6}/3 & 0 & -\sqrt{3}/3 \\
\end{array}
\right)
\end{align}
for the second atom. For convenience, we choose as global spin quantization axis to be the crystallographic $c$-axis, but employ local coordinates for the orbital definitions. The spin operator is modified as $R_1S^{\prime}$ for the first atom and $R_2S^{\prime}$ for the second atom, where $S^{\prime}$ is the standard spin operator. As a result, the spin-orbit coupling operator $\mathcal{H}_{\rm SO} = \lambda \sum_i \mathbf{L}_i\cdot\mathbf{S}_i$ has the specific form:
\begin{align}
\mathcal{H}_{\rm SOC} = \lambda \mathbf{c}_i^\dagger \left(D_1 + D_2 \right) \mathbf{c}_i
\end{align}
where $D_1$ and $D_2$ are the spin-orbital matrices for the first and second atom.

\begin{align}
D_1 = 
\left( 
\begin{array}{cccccc}
0 & iB & -iB & 0 & A & A^* \\
-iB & 0 & iB & -A & 0 & -2iC \\
iB & -iB & 0 & -A^* & 2iC & 0 \\
0 & -A^{*} & -A & 0 & -iB & iB \\
A^{*} & 0 & -2iC & iB & 0 & -iB \\
A & 2iC & 0 & -iB & iB & 0
\end{array}
\right)
\end{align}

\begin{align}
D_2=
\left( 
\begin{array}{cccccc}
0 & iB & -iB & 0 & -A & -A^* \\
-iB & 0 & iB & A & 0 & 2iC \\
iB & -iB & 0 & A^* & -2iC & 0 \\
0 & A^* & A & 0 & -iB & iB \\
-A^* & 0 & 2iC & iB & 0 & -iB \\
-A & -2iC & 0 & -iB & iB & 0
\end{array}
\right)
\end{align}
with $A=\frac{\sqrt{2}}{4}+\frac{\sqrt{6}}{12}i$, $B=\frac{\sqrt{3}}{6}$, $C=\frac{\sqrt{6}}{12}$. 
Within each dimer, the local D$_{3h}$ point group symmetry implies only two types of hopping integrals: diagonal ($t_1$) and off-diagonal ($t_2$) with respect to the $t_{2g}$ orbitals, such that the intradimer hopping is given by:
\begin{align}
\mathcal{H}_{\rm hop} =-& \  \mathbf{c}_1^\dagger  \left( 
\begin{array}{cccccc}
t_1 & t_2 & t_2 & 0 & 0 & 0 \\
t_2 & t_1 & t_2 & 0 & 0 & 0 \\
t_2 & t_2 & t_1 & 0 & 0 & 0 \\
0 & 0 & 0 & t_1 & t_2 & t_2 \\
0 & 0 & 0 & t_2 & t_1 & t_2 \\
0 & 0 & 0 & t_2 & t_2 & t_1
\end{array}
\right) \mathbf{c}_2 + H.c.
\end{align}
Similarly, the crystal field is restricted to a trigonal term ($\Delta$), given by:
\begin{align}
\mathcal{H}_{\rm CF} =-& \ \sum_{i=1}^2 \mathbf{c}_i^\dagger \left( 
\begin{array}{cccccc}
0 & \Delta & \Delta & 0 & 0 & 0 \\
\Delta & 0 & \Delta & 0 & 0 & 0 \\
\Delta & \Delta & 0 & 0 & 0 & 0 \\
0 & 0 & 0 & 0 & \Delta & \Delta \\
0 & 0 & 0 & \Delta & 0 & \Delta \\
0 & 0 & 0 & \Delta & \Delta & 0
\end{array}
\right) \mathbf{c}_i
\end{align}

Considering real materials, we expect $|t_2| = 0.2-0.4$\,eV, based on density-functional theory calculations on several compounds with trivalent B' and M = Ir, Ru using the method described in Ref.~\onlinecite{Foyevtsova2013,winter2016}. For $4d$ elements such as Ru and Rh, we expect $U = 2.5-3.5$\,eV and $\lambda=0.1-0.2$\,eV~\cite{Banerjee2016}. In contrast, for $5d$ elements such as Os and Ir, we expect $U = 1.5-2.0$\,eV and $\lambda = 0.3-0.4$\,eV~\cite{Kim2014}.
Throughout, we use the reduced values $\tilde{U} = U/t_2$, $\tilde{\lambda} =
{\lambda}/t_2$, $\tilde{t_1} = t_1/t_2$, $\tilde{\Delta} = \Delta/t_2$ and fix
the ratio of $J_H/U = 0.18$ for simplicity. 

\subsection{Single-Particle Levels}
\label{sec:single-particle}
We first consider the evolution of the single-particle energy levels assuming each dimer has $D_{3h}$ point group symmetry, as described above. In the absence of SOC, the local trigonal crystal field weakly splits the $t_{2g}$ orbitals into singly degenerate $a_{1g}$ ($\frac{1}{\sqrt{3}}(d_{xy} + d_{xz} + d_{yz})$) and doubly degenerate $e_g$ ($\frac{1}{\sqrt{3}}(d_{xy} + e^{\frac{2\pi i}{3}} d_{xz} + e^{-\frac{2\pi i}{3}} d_{yz})$,  $\frac{1}{\sqrt{3}}(d_{xy} + e^{-\frac{2\pi i}{3}} d_{xz} + e^{\frac{2\pi i}{3}} d_{yz})$). By convention, we label such atomic combinations according to their representation in $D_{3h}$, which describes a single trigonally distorted MO$_6$ octahedron. For $\Delta > 0$, the $a_{1g}$ levels lie below the $e_g$ levels, as depicted in Fig.~\ref{fig:level}. The intradimer hopping is purely diagonal in this atomic basis, with $t_{a_{1g}} = t_1+2t_2$ and $t_{e_{g}} = t_1-t_2$. In the absence of SOC, these hoppings lead to the formation of bonding and anti-bonding combinations of atomic $a_{1g}$ and $e_g$ orbitals. Fig.~\ref{fig:level}(b) shows the ordering of such levels for physically relevant parameters, $t_2 >0$ and $t_2 \gg |t_1|$.

The inclusion of spin-orbit coupling (SOC) leads to the splitting of the
single-particle levels. In the limit of strong SOC ($\lambda \gg \Delta,
t_1,t_2$), the local atomic states are more conveniently described in terms of
doublet $j_{1/2}$  (with the spin, orbital and total effective angular momentum
as $S = \frac{1}{2}, L_{\rm eff} = 1, J_{\rm eff} = \frac{1}{2}$, respectively)
and quadruplet $j_{3/2}$ ($S = \frac{1}{2}, L_{\rm eff} = 1, J_{\rm eff} =
\frac{3}{2}$) levels. Intradimer hopping is not diagonal with respect to the
$j_{1/2}/ j_{3/2}$ character, and therefore leads to {\it both} the mixing of
the $j$-states and the formation of bonding/anti-bonding combinations.
Formally, the mixed spin-orbitals at intermediate $\tilde{\lambda}$ can be
labelled according to their double group representation within D$_{3h}$, which
admits three Kramers doublet representations: $e_{1/2}$, $e_{3/2}$ and
$e_{5/2}$, as shown in Fig.~\ref{fig:level}(c). A character table for these
states is given in the Appendix~\ref{app:d3hdouble}.
Of these single-particle
levels, the $e_{3/2}$ states have pure $j_{3/2}$ character, while the $e_{1/2}$
and $e_{5/2}$ states are mixtures of atomic $j_{1/2}$ and $j_{3/2}$ functions. 
As an example of the relation between the atomic relativistic
$j_{1/2}$ / $j_{3/2}$
basis and
the 3 Kramers doublets, in
Fig.~\ref{fig:level} (d) we display the weights
  $\langle \phi\vert j_{3/2}\rangle^2$ of $j_{3/2}$ states
in terms of the  three Kramers doublets $e_{1/2}$, $e_{3/2}$ and
$e_{5/2}$  ( $\vert\phi\rangle$ states)
as a function of $\tilde{\lambda}$.
  In the non-relativistic limit,
 $\tilde{\lambda} = 0$,
 the 
states are distributed to all relativistic states. With
increasing $\tilde{\lambda}$, the contribution to  $j_{3/2}$ of the lowest 
bonding states $e_{1/2}$  and $e_{3/2}$  arises and finally these states become
pure $j_{3/2}$ while the highest antibonding state $e_{1/2}$
is finally a $j_{1/2}$ state, as it is
the  antibonding $e_{5/2}$. In contrast,
  bonding $e_{5/2}$ shows a non-monotonous composition
  of  $j_{3/2}$  and $j_{1/2}$
 with $\tilde{\lambda}$.

As a function of SOC strength $\tilde{\lambda}$,
the ordering of the single-particle levels changes, such that the ground state
of a single dimer is sensitive to both $\tilde{\lambda}$ and filling. For
intermediate values of $\tilde{\lambda}$ found in real materials, this also has
the effect of confining the $j_{1/2}$ bonding and $j_{3/2}$ antibonding levels
into a narrow energy range, which significantly reduces the energy scales
relevant to the magnetic response of the dimer. The consequences for specific
fillings are considered in sections \ref{sec:d45}--\ref{sec:d25} with inclusion
of Coulomb interactions.

\section{Definition of Effective Moments}
\label{sec:moments}
Given that we are interested in dimers with open-shell (e.g.~doublet) ground states, it is useful to characterize them according to their magnetic response.
For isolated magnetic ions or dimers, it is conventional to describe the molar magnetic susceptibility in terms of temperature-dependent effective moments $\mu_{\rm eff}^\alpha (T)$, defined by:
\begin{align}
\chi^\alpha (T) = \frac{N_A}{3k_BT}[\mu_{\rm eff}^\alpha (T)]^2 \label{eq:effmom}
\end{align}
where $\alpha \in \{x,y,z\}$. In simple cases where there are no low-lying excited multiplets, and the ground state is a pure spin multiplet (there is no orbital degeneracy, and spin-orbit coupling can be neglected), then the susceptibility should follow a pure Curie Law, with $\mu_{\rm eff} = g_S\sqrt{S(S+1)}$ being temperature-independent and isotropic. However, when any of these conditions are violated, as it is common in real materials with strong SOC, the susceptibility of isolated magnetic species will generally be a more complex function of temperature. 

Additional non-Curie contributions to $\chi(T)$ occur, essentially, from two sources. First, there can exist low-lying excited multiplets that become thermally populated at relevant temperatures. Second, when spin-orbit coupling is relevant, it is important to note that the SOC operator $\mathcal{H}_{\rm SOC} = \lambda \mathbf{L} \cdot \mathbf{S}$ generally does not commute with the Zeeman operator, given by $\mathcal{H}_Z = \mu_B\mathbf{H}\cdot \mathbf{M}$, with $\mathbf{M} = (g_S \mathbf{S} + g_L \mathbf{L})$. As a result, the magnetic field induces mixing between different multiplets, introducing additional terms in the susceptibility analogous to the phenomenon of van Vleck paramagnetism. These effects are well documented, and have been studied for isolated ions both theoretically and experimentally~\cite{Kotani1949}. While it is clear that exchange couplings $\sim J$ between magnetic species further impact the paramagnetic susceptibility, provided the temperature is sufficiently large $T\gg J$, then $\chi(T)$ should generally be dominated by local single-species effects. As a result, the behavior of $\mu_{\rm eff}^\alpha (T)$ represents a first clue regarding the nature of the local electronic ground state. We focus on this quantity in the following sections.

\section{ $d^{4.5}$ filling}\label{sec:d45}
\subsection{Survey of Materials}

We first consider the case of $d^{4.5}$ filling that corresponds to three holes
per dimer. This filling can be found in {\abmo} with trivalent B' and M = Rh or
Ir, among which mostly the iridates have been studied experimentally. All of
them show strong deviations from the Curie-Weiss behavior and develop a
characteristic bend in the inverse susceptibility at
$50-100$\,K~\cite{Sakamoto2006}. The linear fit to the high-temperature part
returns the effective moments of 1.53\,$\mu_B$/f.u. (B' = In) and
1.79\,$\mu_B$/f.u. (B' = Sc) that were interpreted as the $S=\frac12$ state of
the mixed-valence dimer~\cite{Sakamoto2006}, although very high Curie-Weiss
temperatures of several hundred Kelvin put into question the validity of such a
fit. Below the bend, a much lower effective moment of 0.76\,$\mu_B$/f.u. (B' =
In) is obtained~\cite{Dey2017}, and a magnetic entropy on the order of $R\ln 2$
is released~\cite{Doi2004,Sakamoto2006,Dey2017}, suggesting the ground-state
doublet of hitherto unknown nature.

At even lower temperatures, interactions between the dimers come into play.
Ba$_3$YIr$_2$O$_9$ undergoes long-range magnetic ordering at
4.5\,K~\cite{Dey2014,panda2015}, whereas Ba$_3$InIr$_2$O$_9$ reveals persistent spin
dynamics down to at least 20\,mK, with local probes suggesting a collective,
possibly spin-liquid behavior of local moments below 1\,K~\cite{Dey2017}.
Ba$_3$ScIr$_2$O$_9$ does not show long-range magnetic order down to at least
2\,K, but pends experimental characterization at lower
temperatures~\cite{Dey2014}.

Rh atoms can also be accommodated in the 6H perovskite structure, but show
strong site mixing with the B' atoms~\cite{Byrne1970,kumar2016}. Therefore, no
experimental information on the magnetism of pure Rh$_2$O$_9$ dimers is
presently available.

\begin{figure}[b]
\includegraphics[angle=0,width=0.95\linewidth]{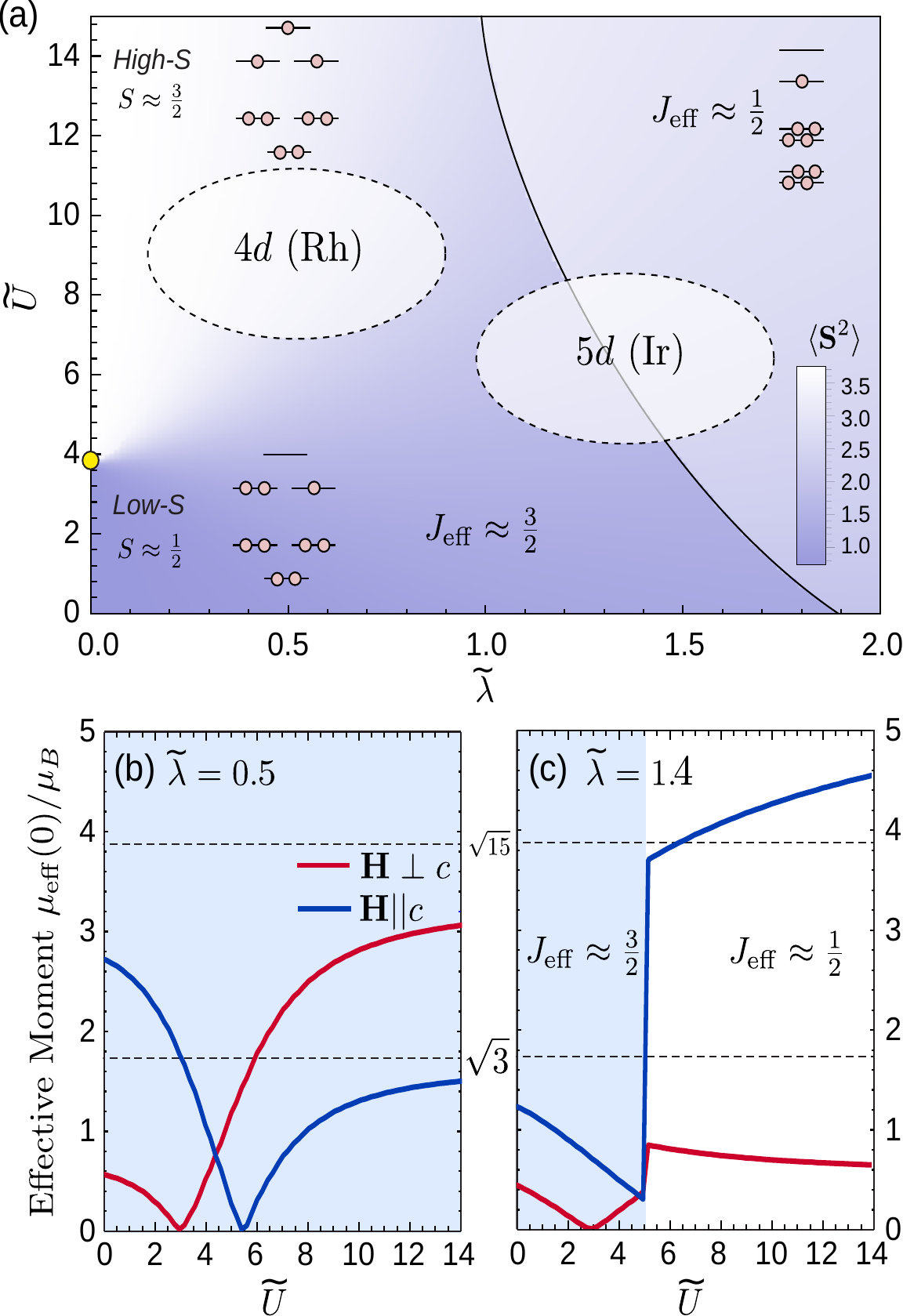}
\caption{(a) Phase diagram of theoretical ground state of a $d^{4.5}$ dimer as a function of $\tilde{U}=U/t_2$ and $\tilde{\lambda}={\lambda}/t_2$ for $t_1$ = $\Delta$ = 0. We fix $J_H/U = 0.18$. (b,c) Evolution of the zero-temperature effective moment (Eq.~\ref{eq:effmom}) for $\tilde{\lambda}$ values corresponding to $4d$
and $5d$ materials, respectively. 
}
\label{fig:d45}
\end{figure}

\begin{figure}[t]
\includegraphics[angle=0,width=0.95\linewidth]{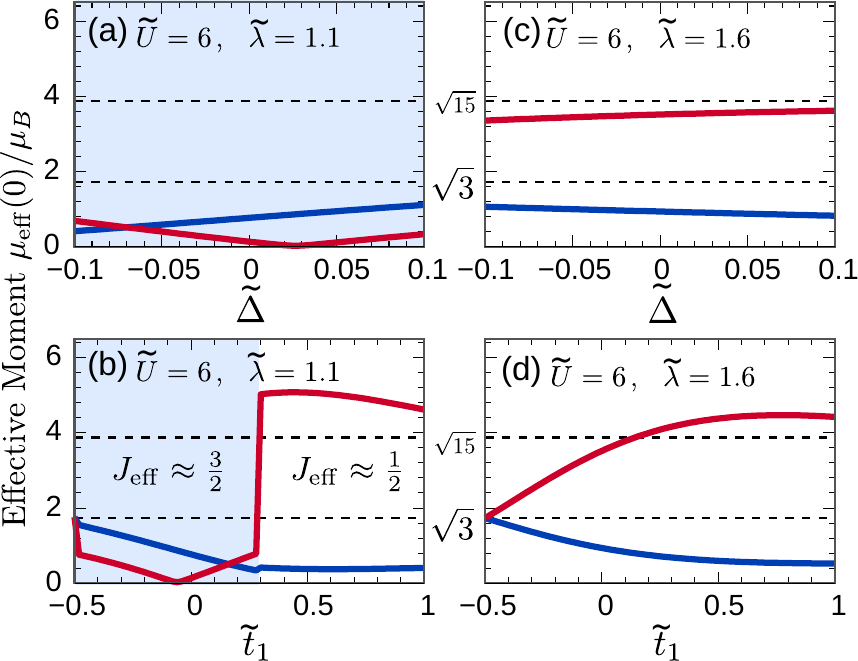}
\caption{Evolution of the zero-temperature effective moment
	(Eq.~\ref{eq:effmom}) 
	as a function of $\tilde{\Delta}=\Delta/t_2$ and $\tilde{t}_1=t_1/t_2$.
}
\label{fig:d45_delta}
\end{figure}

\subsection{Phase Diagram}

The theoretical ground state of a $d^{4.5}$ dimer as a function of $\tilde{U}$ and
$\tilde{\lambda}$ is shown in Fig.~\ref{fig:d45}(a) for $t_1 = \Delta = 0$. The color shading indicates the expectation value of the total spin moment squared $\langle \hat{\mathbf{S}}^2 \rangle$ per dimer. 
In the
non-relativistic limit $\tilde{\lambda} \to 0$, there are two possible ground states
depending on the relative strength of $\tilde{t}_1$ and $\tilde{U}$ (and $J_H/t_2$
since we fix the ratio $J_H/U =0.18$). If the
Hund's coupling is large compared to the splitting of the anti-bonding $e_g$
and $a_{1g}$ levels ($J_H \gg t_2$), the ground state is a high-spin
$S=3/2$ quartet with nominally one hole in each of the anti-bonding $e_{3/2}$,
and $e_{5/2}$ orbitals (Fig.~\ref{fig:d45} (a) upper left corner).
In this configuration, the orbital angular momentum is completely quenched, and $\langle \hat{\mathbf{S}}^2\rangle = S(S+1) = 15/4$. In contrast, for $J_H \ll t_2$, a low-spin $S=1/2$ configuration is
instead preferred, with two holes nominally occupying the $a_{1g}$ anti-bonding
state, and one hole occupying the $e_g$ anti-bonding state
(lower left corner in Fig.~\ref{fig:d45} (a)). This is indicated by $\langle \hat{\mathbf{S}}^2\rangle = 3/4$. Provided $C_3$
rotational symmetry is preserved, this latter configuration has unquenched orbital
angular momentum, as the $e_g$ hole may occupy either the $e_{3/2}$ or
$e_{5/2}$ single-particle levels shown in Fig.~\ref{fig:level}(c). The combined spin
and orbital degrees of freedom provide an overall four-fold ground-state
degeneracy for $\tilde{\lambda} = 0$. As the partially occupied single-particle levels are predominantly of $j_{3/2}$ character, we refer to this ground state as having total $J_{\rm eff} \approx 3/2$. 

For both the high-spin $S \approx 3/2$ and low-spin $J_{\rm eff} \approx 3/2$ cases, introducing small $\tilde{\lambda}$ leads to
a splitting of the quartet states into pairs of Kramers doublets. In the
high-spin limit, this can be viewed as the introduction of an on-site
zero-field splitting term $(S_x^2+S_y^2+S_z^2)$ that energetically prefers the
$m_S = \pm 1/2$ states. In the low-spin limit, SOC leads to spin-orbital
locking of the unquenched $L$ and $S$ moments, stabilizing the $m_J = \pm 1/2$
states.  With finite $\tilde{\lambda}$, these cases are smoothly connected, such that
there is no sharp
``phase transition''  for small $\tilde{\lambda}$ on increasing $\tilde{U}$. As indicated in Fig.~\ref{fig:d45}(a), $\langle \hat{\mathbf{S}}^2\rangle$ evolves continuously between the high-spin and low-spin limits, as $S$ is no longer a good quantum number with finite SOC.

Further increasing $\tilde{\lambda}$ leads to a change in the ordering of the single-particle $e_{1/2}$ bonding and $e_{5/2}$ anti-bonding levels, as discussed in Section~\ref{sec:single-particle}. This stabilizes a different ``low-spin'' $J_{\rm eff} \approx {1/2}$ ground state, with nominally two holes in the anti-bonding $j_{1/2}$ level ($e_{5/2}$), and one hole in the bonding $j_{1/2}$ level ($e_{1/2}$). 

Considering real materials, hypothetical Rh-based dimers should fall well
within the high-spin $S \approx 3/2$ region. In contrast, Ir-based dimers may
lie on the border of the $J_{\rm eff} \approx 1/2$ and $J_{\rm eff} \approx
3/2$ regions if microscopic parameters determined by resonant inelastic x-ray
scattering (RIXS) on integer-valence Ir$_2$O$_9$
dimers~\cite{Revelli2019,nag2019hopping} are used. Similar experiments on the
mixed-valence compounds would be required to identify the experimental ground
states of real materials. In this context, a useful observation is that the
different electronic configurations of the dimers can also be distinguished by
the anisotropy and magnitude of the effective magnetic moments $\mu_{\rm
eff}^\alpha (T)$, as defined in Section \ref{sec:moments}. 

In Fig.~\ref{fig:d45}(b) we show the evolution of the zero-temperature limit
of the effective moment per dimer $\mu_{\rm eff}(0)$ as a function of
$\tilde{U}$ for $\tilde{\lambda} = 0.5$, corresponding
to the expected range for $4d$ dimers of Rh. In this case, SOC reduces
the average value of $\mu_{\rm eff}(0)$ to well below the spin-only value of
$\sqrt{15}$ for pure $S=3/2$ moments for all values of $\tilde{U}$. The
crossover from the low-spin to high-spin ground state with increasing
$\tilde{U}$ leads to a reversal of the anisotropy of the effective moment; for
small $\tilde{U}$, $\mu_{\rm eff}^{ab}<\mu_{\rm eff}^c$, while for large
$\tilde{U}$, $\mu_{\rm eff}^{ab}>\mu_{\rm eff}^c$. We expect that hypothetical
Rh-based dimers should fall into the latter category.

In the region of parameters applicable to Ir-based dimers ($\tilde{\lambda}=1.4$; Fig.~\ref{fig:d45}(c)), increasing $\tilde{U}$ leads to a phase transition from
a $J_{\rm eff} \approx 3/2$ to $J_{\rm eff} \approx 1/2$ ground state, which is marked by a significant increase of the average effective moment. The average moment in the $J_{\rm eff} \approx 3/2$ state is strongly suppressed by SOC, and displays an anisotropy that is strongly sensitive to various parameters such as $\tilde{\lambda}, \tilde{U}, \tilde{\Delta}$, and $\tilde{t}_1$. The influence of the latter two parameters is shown in Fig.~\ref{fig:d45_delta}(a,b). Modifying $\tilde{\Delta}$ and $\tilde{t}_1$ can lead to reversals of the anisotropy of the effective moment in the $J_{\rm eff} \approx 3/2$ region. Greatly increasing $\tilde{t}_1$ also stabilizes the $J_{\rm eff} \approx 1/2$ ground state, ultimately driving the phase transition. The $J_{\rm eff} \approx 1/2$ ground state at large $\tilde{U}$ and/or $\tilde{t}_1$ is characterized by a much larger average moment, which is strongly anisotropic with $\mu_{\rm eff}^c > \mu_{\rm eff}^{ab}$. As shown in Fig.~\ref{fig:d45_delta}(c,d), this anisotropy is only weakly affected by crystal field $\tilde{\Delta}$, and remains for a wide range of values of $\tilde{t}_1$ hopping. 

\subsection{Comparison to Experiment}

\begin{figure}[t]
\includegraphics[angle=0,width=\linewidth]{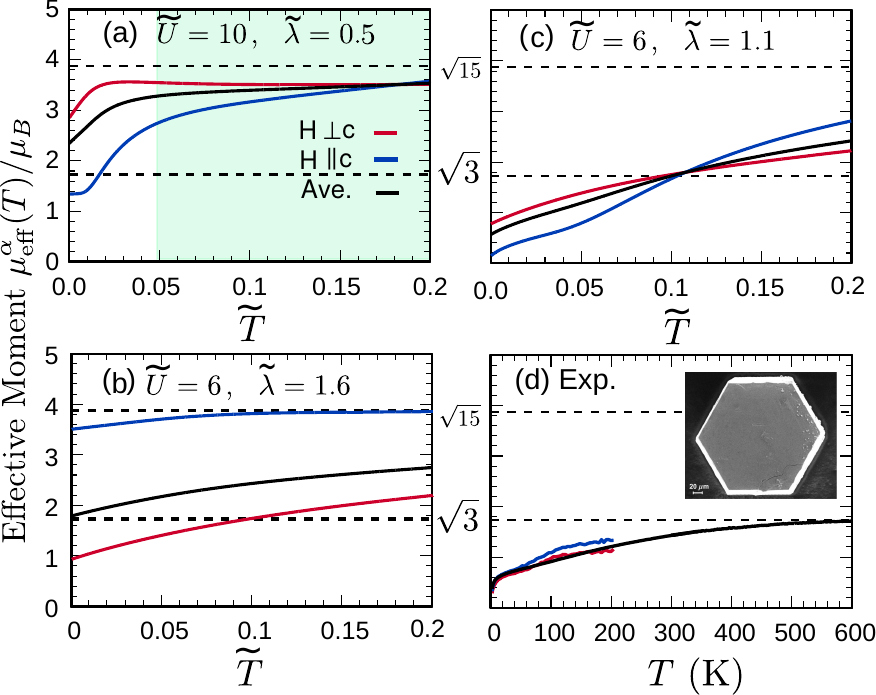}
\caption{Temperature dependence of the magnetic moments for $d^{4.5}$ dimers
	for various $\tilde{U} (U/t_2)$ and $\tilde{\lambda}({\lambda}/t_2)$
	values for $t_1$ = $\Delta$ = 0. (a) Parameters relevant for isolated
	$4d$ dimers. The green area displays the region where
	the first excited state is occupied.
	(b,c) Parameters relevant for isolated $5d$ dimers in $J_{\rm eff}
	\approx 1/2$ and $J_{\rm eff} \approx 3/2$ ground states, respectively.
	The energy of the first excited state is above 0.2 $t_2$. (d)
	Experimental effective moment for Ba$_3$InIr$_2$O$_9$ derived from both
	single-crystal and powder measurements. The insert of figure 4 (d) is
	the scanning electron microscopy image of the Ba$_3$InIr$_2$O$_9$
	crystal.} \label{fig:d45T}
\end{figure}

For temperatures large compared to the interactions between dimers, the effective moment per dimer can be extracted from experimental susceptibility
 data $\chi^{\alpha}$ via:
 \begin{align}
\mu^{\alpha}_{\rm eff}(T) \approx \sqrt{\frac{3k_BT}{N_A}\chi^{\alpha}_{\rm exp}(T)}
\label{eq:mu}
\end{align}
where $\alpha$ indicates the field direction,
$N_A$ is the Avogadro constant, $k_B$ is the Boltzmann constant, and $T$ is the temperature.
Given that all Ir-based $d^{4.5}$ dimers discussed above show features in the specific heat at low temperatures $T \lesssim 10$\,K that account for a significant fraction of $R\ln 2$ entropy, we assume that the magnetic couplings between dimers are relatively weak. As a result, the temperature dependence of $\chi(T)$ above room temperature should be largely dominated by the evolution of $\mu_{\rm eff}(T)$. In Fig.~\ref{fig:d45T}(a)--(c), we thus show the theoretical temperature dependence of the effective moment for selected parameters corresponding to regions expected for real materials. For real materials, $|t_2|=0.2-0.4$\,eV, such that room temperature corresponds to roughly $\tilde{T} = T/t_2 \sim 0.1$.

For hypothetical Rh-based dimers (Fig.~\ref{fig:d45T}(a)), we find that thermal fluctuations rapidly overcome the  ``single-dimer'' spin-anisotropy from SOC, restoring the spin-only value of $\sqrt{15} \ \mu_B$ as temperature is raised. Nevertheless, below room temperature such materials will display strong deviations from Curie behavior even in the absence of interdimer couplings. The green area above 0.047 is the region 
where the first excited state is occupied.

For Ir-based dimers, we consider $\tilde{\lambda} = $ 1.1 and 1.6 corresponding to the $J_{\rm eff} \approx 3/2$ and $J_{\rm eff}\approx 1/2$ ground states, respectively (Fig.~\ref{fig:d45T}(b,c)). In both cases, $\mu_{\rm eff}(T)$ is a monotonically increasing function. The energy of the first excited state is above 0.2 $t_2$. For the $J_{\rm eff}\approx 1/2$ ground state (Fig.~\ref{fig:d45T}(c)),
the effective moment is strongly anisotropic, with an average value that
exceeds the value of $\sqrt{3}$ expected for pure $J_{\rm eff} = 1/2$ states. A
similar effect is known for isolated ions with $J_{\rm eff} = 1/2$ ground
states~\cite{Kotani1949}, and therefore is not unique to the dimer case. In
contrast, the anisotropy is expected to be weaker in the $J_{\rm eff} = 3/2$
ground state (Fig.~\ref{fig:d45T}(b)), with average values below $\sqrt{3}$ at experimentally relevant
temperatures. 

Our results suggest that the presence or absence of this anisotropy can be
used to diagnose  the electronic ground state of mixed-valence dimers.
Unfortunately, this anisotropy aspect remains unexplored, as most of the
experimental data reported so far have been measured on powders, with no single
crystals available. We could overcome this problem by growing a small single
crystal of Ba$_3$InIr$_2$O$_9$ (see Appendix~\ref{app:prep} for details of
crystal growth and characterization) and measured its magnetic susceptibility
along different directions. Data up to 200\,K were obtained, while at
higher temperatures the signal drops below the sensitivity threshold of the
magnetometer, but already these data indicate weak anisotropy of the
paramagnetic response. The effective moment is quite low and approaches $\sqrt
3$ at elevated temperatures. These observations strongly suggest that the
ground state of the mixed-valence dimer in Ba$_3$InIr$_2$O$_9$ is $J_{\rm eff}
= 3/2$. A similar $J_{\rm eff} = 3/2$ scenario can be envisaged for other
mixed-valence iridates with trivalent B', where the susceptibility data around
room temperature also yield effective moments of about $\sqrt
3$~\cite{Sakamoto2006}.

An ideal $J=3/2$ state is a quartet with zero magnetic dipole moment, but under
experimental conditions it will usually split into two Kramers doublets, which
is also the case here at $\tilde{\lambda}\neq 0$. This explains why at low
temperatures magnetic entropy of not more than $R\ln 2$ is
recovered~\cite{Doi2004,Dey2014,Dey2017}, corresponding to the lower Kramers
doublet. Low-temperature magnetism associated with this doublet should be then
treated as an effective spin-$\frac12$ with a strongly renormalized $g$-tensor.
The very low $g$-value elucidates the fact that magnetization of
Ba$_3$InIr$_2$O$_9$ does not reach saturation even at 14\,T, even though
exchange couplings are on the order of several Kelvin and would be easily
saturated in a conventional spin-$\frac12$ antiferromagnet with
$g=2.0$~\cite{Dey2017}.

\section{ $d^{3.5}$ filling}\label{sec:d35}
\subsection{Survey of Materials}

Materials with $d^{3.5}$ filling are found for two different compositions {\abmo}. The first case corresponds to M being a group 8 element (Os, Ru) and trivalent B', whereas the second case corresponds to M being a group 9 element (Ir, Rh) and monovalent B'. 

The mixed-valence ruthenates have been reported for a wide range of trivalent
B' ions, including Y, In, and many of the lanthanides. All these compounds show
strong deviations from the Curie-Weiss
behavior~\cite{DOI2001,DOI2002,SHLYK2007}. An effective spin-$\frac12$ state
was conjectured~\cite{Ziat2017} for Ba$_3$B'Ru$_2$O$_9$ with B' = In, Y, Lu
from the relatively low values of the magnetic susceptibility and from the
local moment of 1.0\,$\mu_B$/dimer in the magnetically ordered state of
Ba$_3$YRu$_2$O$_9$~\cite{Senn2013}. On the other hand, Ba$_3$LaRu$_2$O$_9$
shows an overall higher susceptibility that approaches the Curie-Weiss regime
for spin-$\frac32$ above room temperature~\cite{chen2020}. The local moment of
$2.6-2.8$\,$\mu_B$/dimer in the magnetically ordered
state~\cite{Senn2013,chen2020} also supports the spin-$\frac32$ scenario. These
dissimilar trends indicate the presence of at least two competing electronic
states in the $d^{3.5}$ Ru$_2$O$_9$ dimers. Indeed, local excitations of
magnetic origin were observed by inelastic neutron scattering around 35\,meV
and assigned to a transition between these states tentatively identified as
$S=1/2$ and $S=3/2$ states of the mixed-valence dimer~\cite{Ziat2017,chen2020}.

Similar to the $d^{4.5}$ iridates, the $d^{3.5}$ ruthenates exhibit a range of
magnetic ground states, from long-range order confirmed in B' = La, Nd,
Y~\cite{Senn2013,chen2020,DOI2001} and anticipated for B' = Lu~\cite{Ziat2017},
to a static disordered state in Ba$_3$InRu$_2$O$_9$~\cite{Ziat2017}. To our
knowledge, Os-based dimers with $d^{3.5}$ filling have yet to be explored.

Examples of the second case of materials include
Ba$_3$(Na/Li)Ir$_2$O$_9$~\cite{Loye2009,KIM2004}. Susceptibility measurements
show signs of magnetic order at 75 and 50\,K for B' = Li and Na,
respectively~\cite{KIM2004}. Curie-Weiss fits of the magnetic susceptibility
between about 150\,K and room temperature return the effective moments of
3.93\,$\mu_B$/f.u. (B' = Li) and 3.60\,$\mu_B$/f.u. (B' = Na)~\cite{KIM2004}.
On the other hand, Curie-Weiss temperatures of, respectively, $-576$\,K and
$-232$\,K suggest that a true paramagnetic regime may not have been reached in
this temperature range. To our knowledge, Rh-based dimers with $d^{3.5}$
filling were not reported. 

\subsection{Phase Diagram}

\begin{figure}[t]
\includegraphics[angle=0,width=\linewidth]{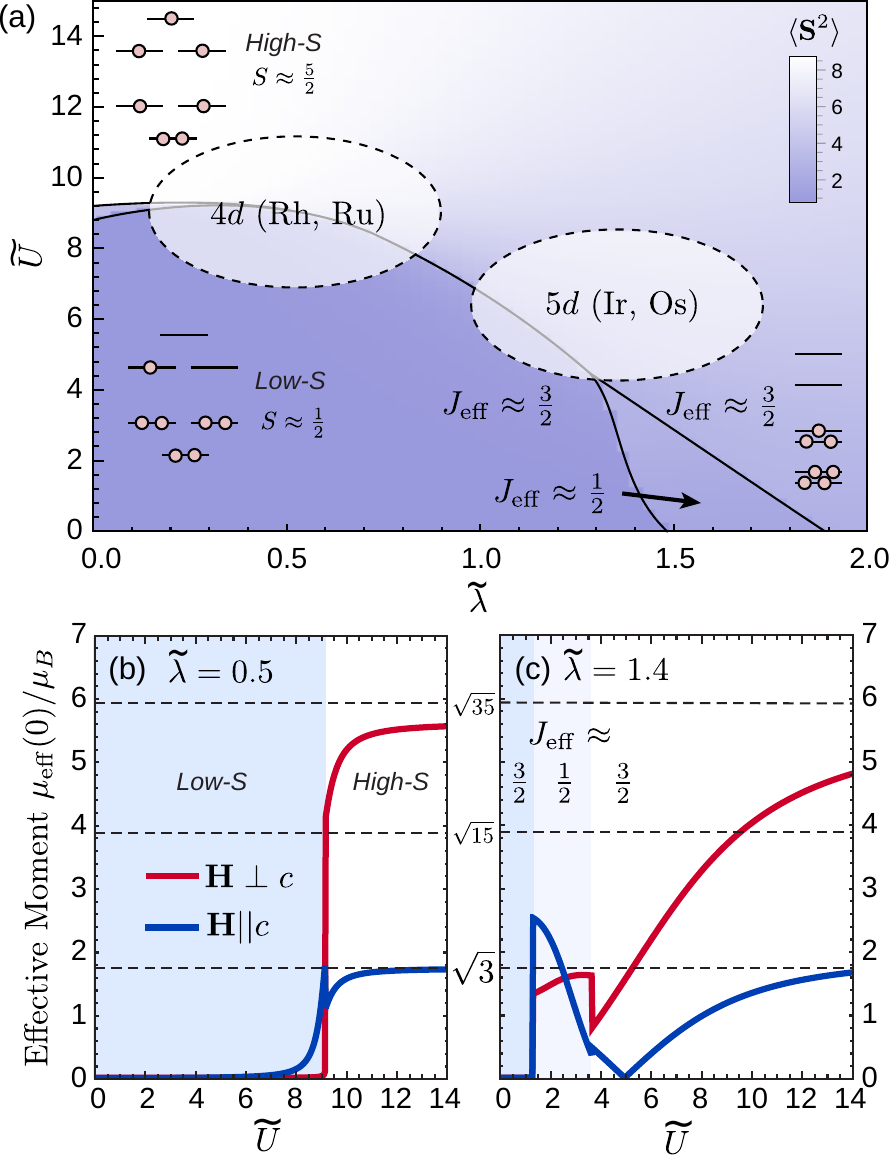}
\caption{(a) Phase diagram of theoretical ground state of a $d^{3.5}$ dimer as
	a function of $\tilde{U}=U/t_2$ and $\tilde{\lambda}={\lambda}/t_2$ for $t_1$ = $\Delta$ = 0. We
	fix $J_H/U = 0.18$. (b,c) Evolution of the zero-temperature effective
	moment as a function of $\tilde{U}$ for $\tilde{\lambda}$ values corresponding
	to $4d$ and $5d$ materials, respectively.
	Note that the jumps in the figures are due to changes in
	$J_{\rm eff}$.}  
 \label{fig:d35}
\end{figure}

The theoretical ground state of a $d^{3.5}$ dimer as a function of $\tilde{U}$ and $\tilde{\lambda}$ is shown in Fig.~\ref{fig:d35}(a) for $t_1 = \Delta = 0$. Similar to the $d^{4.5}$ case, the non-relativistic limit $\tilde{\lambda} \to 0$ features high-spin and low-spin states, which are stabilized for large and small $J_H/t_2$, respectively. Large Hund's coupling produces a high-spin $S=5/2$ state with nominally one hole in each of the anti-bonding $e_{3/2}$, $e_{5/2}$ and bonding $e_{1/2}, \ e_{3/2}$ orbitals. In this state, the orbital angular momentum is essentially quenched. In contrast, weak interactions lead to a low-spin $S=1/2$ ground state, with two holes nominally occupying the $a_{1g}$ anti-bonding level, and three holes occupying the $e_g$ anti-bonding levels. This low-spin ground state has orbital degeneracy associated with the single electron in the $e_g$ anti-bonding levels implying an unquenched orbital moment. In the non-relativistic limit, the ground state is four-fold degenerate, with unpaired electrons occupying single-particle levels with predominant $j_{3/2}$ character. As a result, we refer to this low-spin ground state as $J_{\rm eff} \approx 3/2$. A very narrow intermediate-spin $S \approx 3/2$ phase may also lie between these cases for weak SOC.

In contrast to the $d^{4.5}$ case, a distinct transition occurs between the low-spin and high-spin states, as can be seen in $\mu_{\rm eff}(0)$ as a function of $\tilde{U}$, shown in Fig.~\ref{fig:d35} (b) for $\tilde{\lambda} = 0.5$. The introduction of small $\tilde{\lambda}$ leads to spin-orbital locking in the low-spin phase, which ultimately quenches completely the effective moment. As a result, $\mu_{\rm eff}(0) \approx 0$ for small $\tilde{U}$. From the perspective of the single-particle levels, the unpaired electron occupies an $e_{3/2}$ level, which has pure $j_{3/2}$ character. As a result, the effective moment is precisely zero. For the high-spin case, small $\lambda$ leads to a splitting of the six degenerate $S=5/2$ states into three pairs of Kramers doublets. The $m_S = \pm 1/2$ doublet forms the ground state, which leads to $\mu_{\rm eff}^{ab}>\mu_{\rm eff}^c \approx \sqrt{3}$. Considering real materials, we anticipate that Rh- and Ru-based dimers may lie in an interesting region on the border of the high-spin and low-spin states.

Returning back to the phase diagram Fig.~\ref{fig:d35}(a), we can also consider
the sequence of ground states that occur for weak interactions as a function of
$\tilde{\lambda}$. These can be understood in the single particle picture from
Fig.~\ref{fig:level}(c). As noted above, for small $\tilde{\lambda}$, the $J_{\rm eff} \approx 3/2$ ground state
has a single electron occupying an $e_g$ anti-bonding level of $e_{3/2}$
symmetry, with pure $j_{3/2}$ anti-bonding character.  With increasing
$\tilde{\lambda}$, this level crosses the $j_{1/2}$ bonding level of $e_{1/2}$
symmetry, leading instead to a ground state with $J_{\rm eff} \approx 1/2$
nature. Finally, further increasing $\tilde{\lambda}$ leads to another crossing of the
$e_{1/2}$ level with an $e_{5/2}$ level that is nominally $j_{3/2}$
anti-bonding. As a result, the ground state in the $\tilde{\lambda} \to \infty$ limit
consists of two holes in the anti-bonding $j_{1/2}$ level ($e_{5/2}$), two
holes in the bonding $j_{1/2}$ level ($e_{1/2}$) and one hole in the
antibonding $j_{3/2}$ level ($e_{5/2}$). This latter $J_{\rm eff} \approx 3/2$
ground state is smoothly connected to the high-spin $S=5/2$ state appearing at
weak SOC.

In Fig.~\ref{fig:d35}(c), we show the evolution
of the zero-temperature effective moment as a function of $\tilde{U}$ for $\tilde{\lambda} = 1.4$, corresponding to $5d$ materials.
In this case, $\mu_{\rm eff}(0)$ can be used to distinguish the three different ground states. For small $\tilde{U}$, the effective moment is 0, as the low-energy degrees of freedom have pure $J_{\rm eff} \approx 3/2$ character. When entering into the $J_{\rm eff} \approx 1/2$ state upon increasing $\tilde{U}$, the effective moment jumps to an average value $\sim \sqrt{3}\mu_B$. With further increasing $U$, the average effective moment again drops when entering the $J_{\rm eff} \approx 3/2$ ground state. However, interactions ultimately enhance the spin contribution to the moment, such that $\mu_{\rm eff}$ tends to grow with increasing $U, J_H$. The anisotropy $\mu_{\rm eff}^{ab} > \mu_{\rm eff}^c$ is maintained throughout this latter $J_{\rm eff} \approx 3/2$ phase. 
 
\subsection{Comparison to Experiment}

\begin{figure}[t]
\includegraphics[width=\linewidth]{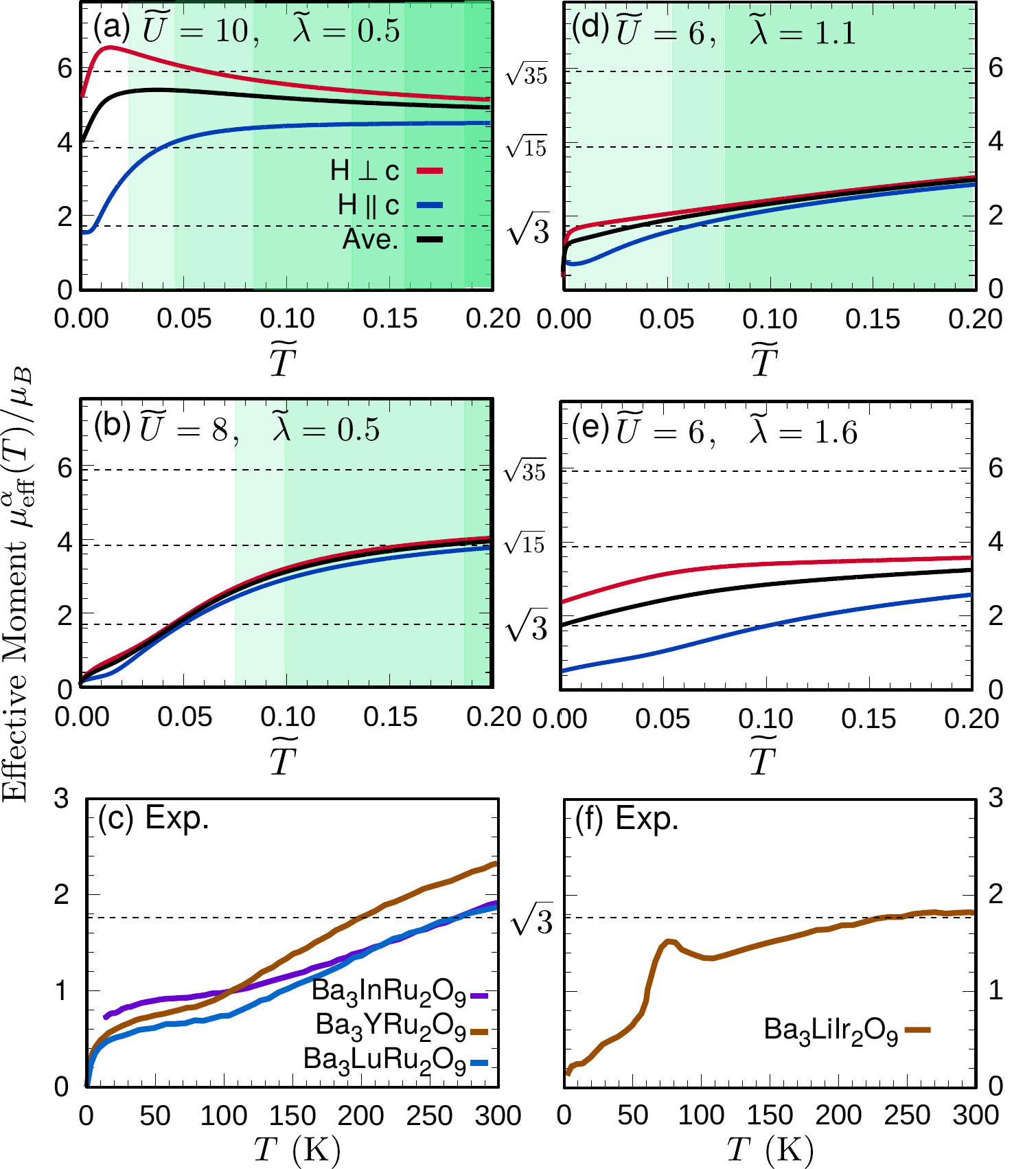}
\caption{Temperature dependence of the magnetic moments for $d^{3.5}$ dimers
	with various $\tilde{U}$ and $\tilde{\lambda}$ values for $t_1$ =
	$\Delta$ = 0. (a,b) Theoretical results for isolated $4d$ dimers in
	high-spin and low-spin ground states, respectively. The green areas
	display the regions where the 6 lowest excited states are filled
	sequentially. (c) Experimental effective moment for various Ru-based
	dimers based on data from Ref.~\onlinecite{Ziat2017}. 
(d,e) Theoretical results for isolated $5d$ dimers in different $J_{\rm eff}
	\approx 3/2$ ground states. The green areas in (d) 
	display the regions where the 3 lowest excited states are filled
        sequentially. The energy of the first
	excited state in (e) is
	above 0.2 $t_2$. (f) Experimental effective moment for
	Ba$_3$LiIr$_2$O$_9$ derived from Ref.~\onlinecite{KIM2004}.}
	\label{fig:d35T}
\end{figure}

In Fig.~\ref{fig:d35T}, we show the temperature dependence of the effective moment for selected parameters corresponding to regions expected for real materials. For Ru/Rh systems, we consider $\tilde{\lambda} = 0.5$, with $\Delta = t_1 = 0$ for simplicity. Fig.~\ref{fig:d35T}(a) and (b) show the expected behavior for the high-spin and low-spin ground states, respectively. The green areas display the regions
where the lowest  excited states are sequentially filled as a function of temperature. 
For the high-spin state, we find a strong temperature dependence as thermal fluctuations rapidly overwhelm the local ``single-ion'' magnetic anisotropies. Thus the anisotropy of the effective moment is suppressed with increasing $T$, and the average value becomes comparable to the pure spin value of $\sqrt{35}\mu_B$ for $S=5/2$. For the low-spin state, $\mu_{\rm eff}^\alpha(T)$ is nearly isotropic at all temperatures, and slowly decreases to zero as the temperature is lowered. Considering real materials, Fig.~\ref{fig:d35T}(b) closely resembles the response of the powder samples of Ba$_3$YRu$_2$O$_9$, Ba$_3$LuRu$_2$O$_9$, and Ba$_3$InRu$_2$O$_9$ reported in Refs.~\onlinecite{DOI2002,Ziat2017}. 
Moreover, magnetic susceptibility measured on a single crystal of
Ba$_3$InRu$_2$O$_9$ shows a nearly isotropic magnetic
response~\cite{SHLYK2007}, in agreement with the low-spin scenario. In
Fig.~\ref{fig:d35T}(c), we plot the $\sqrt{(3k_BT/N_A)\chi_{\rm avg}(T)}$ for
all three compounds, derived from the experimental $\chi(T)$ in
Ref.~\onlinecite{Ziat2017}. As above, we expect $|t_2|=0.2-0.3$\,eV, so room
temperature corresponds to roughly $\tilde{T} = T/t_2 \sim 0.1$. 

Given this correspondence, we can conclude that the Ru-based dimers very likely
have low-spin ground states. Similar to the $d^{4.5}$ iridates, this $J_{\rm
eff}=3/2$ state is represented by two Kramers doublets. We emphasize that the
suppression of $\mu_{\rm eff}(T)$ at low temperatures results from SOC, which
couples the $S=1/2$ moment per dimer with the unquenched orbital moment such
that the total effective moment is nearly canceled in the ground state. This
scenario is also compatible with the relatively small ordered moments deduced
from neutron scattering~\cite{Senn2013}, and with the observed electronic
excitation around 35\,meV~\cite{Ziat2017} that should be assigned to the
transition between the two doublets.  

An interesting exception to this picture is the behavior of Ba$_3$LuRu$_2$O$_9$, which shows a larger average $\mu_{\rm eff}$ over an extended temperature range. It may be possible that this material adopts a high-spin or intermediate-spin ground state instead; a non-trivial test of this scenario would be demonstration of significant anisotropy of $\mu_{\rm eff}^\alpha(T)$, with $\mu_{\rm eff}^{ab} > \mu_{\rm eff}^c$, if single crystal samples become available.

For Ir- and Os-based dimers, the two realistic ground states are both described
as $J_{\rm eff} \approx 3/2$. As a result, they are less distinguishable on the
basis of magnetic susceptibility. In Fig.~\ref{fig:d35T} (d,e), we compare
$\mu_{\rm eff}(T)$ for $\tilde{U} = 6$ and different values of $\tilde{\lambda}
= 1.1, \ 1.6$ corresponding the two ground states. The green areas
in In Fig.~\ref{fig:d35T}  (d) display the regions where the
lowest three excited states are filled as a function of temperature.
The energy of the first excited state is
above 0.2 $t_2$ in (e). In both cases, we find $\mu_{\rm eff}^{ab} > \mu_{\rm
eff}^c$, with average values falling in similar ranges. In Fig.~\ref{fig:d35T}
(f), we plot $\sqrt{(3k_BT/N_A)\chi_{\rm avg}(T)}$ for Ba$_3$LiIr$_2$O$_9$,
derived from the experimental susceptibility from Ref.~\onlinecite{KIM2004}.
The experimental value of $\sim 2 \mu_B$ around room temperature is compatible
with the theoretical results on panel (d) at $\tilde{T}=0.1$. On the other
hand, the effective moments of 3.60 and 3.93\,$\mu_B$/f.u. obtained from the
Curie-Weiss fits~\cite{KIM2004} are comparable to the high-temperature limit on
panel (e). A general problem here is that relatively high magnetic ordering
temperatures of 50 and 75\,K indicate sizable exchange interactions between the
dimers. Therefore, it may be difficult to directly compare the theoretical and
experimental data within the available temperature range. 

\section{ $d^{2.5}$ filling}\label{sec:d25}

\subsection{Survey of Materials}
Materials with $d^{2.5}$ filling have {\abmo} compositions with monovalent B'
and transition metal M belonging to group 8 (Os, Ru). Of these,
Ba$_3$NaRu$_2$O$_9$ shows a rather abrupt decrease~\cite{Stitzer2002} in the
magnetic susceptibility at 210\,K, which is accompanied by a structural
transition and interdimer charge order~\cite{Kimber2012}. The low-temperature
phase is thus composed of distinct (Ru$^{6+}$)$_2$O$_9$ and
(Ru$^{5+}$)$_2$O$_9$ dimers, which do not maintain $d^{2.5}$ filling.
Interestingly, the Li analogue also exhibits a drop in susceptibility around
150\,K, although this is not associated with a structural
transition~\cite{Stitzer2002}; it is presently unclear whether charge-order
occurs in this case. The Os analogues, Ba$_3$B'Os$_2$O$_9$ (B' = Na, Li), have
also been reported~\cite{Stitzer2003}. Both compounds show no evidence of
charge order, whereas kinks in the susceptibility around $10-13$\,K likely
indicate magnetic ordering. Curie-Weiss fits yield larger effective moments of
about 5\,$\mu_B$ for the ruthenates~\cite{Stitzer2002} and lower effective
moments of about 3.3\,$\mu_B$ for the osmates~\cite{Stitzer2003}. However, all
values should be taken with caution, because, similar to the $d^{3.5}$
iridates, large antiferromagnetic Weiss constants suggest that the true
paramagnetic regime has not been reached, and/or the effective moment is
strongly temperature-dependent.

\subsection{Phase Diagram}

\begin{figure}[t]
\includegraphics[angle=0,width=\linewidth]{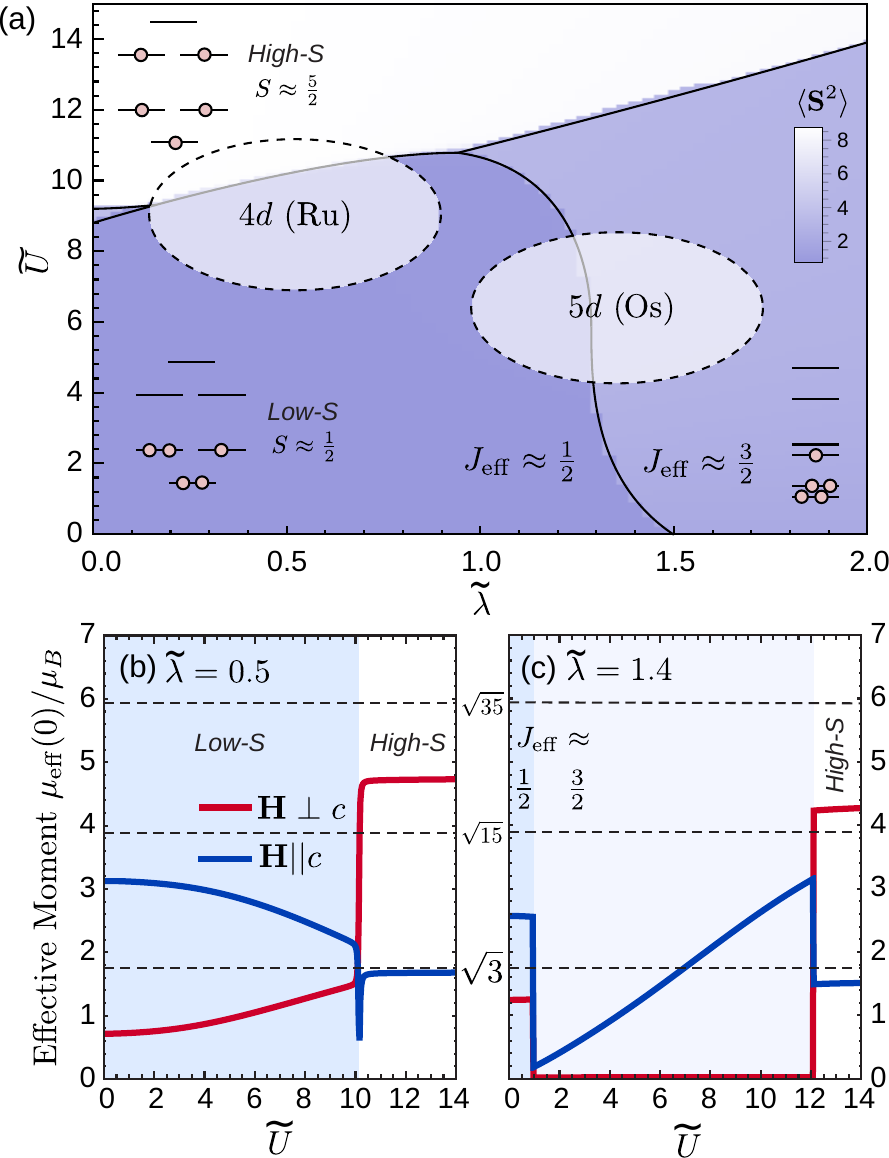}
\caption{P(a) Phase diagram of theoretical ground state of a $d^{2.5}$ dimer as a function of $\tilde{U}=U/t_2$ and $\tilde{\lambda}={\lambda}/t_2$ for $t_1$ = $\Delta$ = 0. We fix $J_H/U = 0.18$. (b,c) Evolution of the zero-temperature effective moment as a function of $\tilde{U}$ for $\tilde{\lambda}$ values corresponding to $4d$ and $5d$ materials, respectively.} 
\label{fig:d25}
\end{figure}

The theoretical phase diagram for $d^{2.5}$ filling and $\Delta = t_1 = 0$ is shown in Fig.~\ref{fig:d25}(a). The green areas display the lowest five excited states region. In the non-relativistic limit $\tilde{\lambda} \to 0$, these materials also exhibit high-spin and low-spin states depending on the strength of interactions. 
For the high-spin $S=5/2$ state, there is one electron in each anti-bonding $e_g$ ($e_{3/2}$, $e_{5/2}$) and bonding $e_g$ ($e_{1/2}, e_{3/2}$) and bonding $a_g$ ($e_{1/2}$) orbitals. At lowest order, the orbital momentum is quenched. In contrast, the low-spin state consists of all five electrons in the bonding orbitals, while the anti-bonding levels are empty. The degenerate bonding $e_g$ levels contain three electrons, leading to an unquenched orbital degree of freedom. However, in this case, the unpaired electron nominally occupies a level with $j_{1/2}$ character, identifying this ground state as $J_{\rm eff} \approx 1/2$. 

The evolution of $\mu_{\rm eff}(0)$ as a function of $\tilde{U}$ is shown in Fig.~\ref{fig:d25}(b) for $\tilde{\lambda} = 0.5$, corresponding to a reasonable value for Ru-based dimers.The energy of the first excited state is above 0.2 $t_2$. For the high-spin state, the introduction of a small $\lambda$ leads to a splitting of the six degenerate states into three pairs of Kramers doublets. The $m_s = \pm 1/2$ states form the ground state, which leads to $\mu_{\rm eff}^{ab} > \mu_{\rm eff}^c \approx \sqrt{3}$. In contrast, the low-spin state has a reversed anisotropy of the effective moment. This finding can be understood from the single-particle levels. For small $\lambda$, the single unpaired electron occupies an $e_{1/2}$ orbital with mostly $j_{1/2}$ character, leading to a $J_{\rm eff} \approx 1/2$ ground state. The average effective moment is then $\approx \sqrt{3}$, with an anisotropy $\mu_{\rm eff}^{ab} < \mu_{\rm eff}^c$ due to mixing with $J_{\rm eff} \approx 3/2$ states.  

Considering the ordering of single-particle levels also provides a description of the third phase appearing at large $\tilde{\lambda}$ in the phase diagram. For small 
$\tilde{U}$ and large $\tilde{\lambda}$, a single unpaired electron occupies an $e_{3/2}$ anti-bonding level of pure $j_{3/2}$ character, suggesting a $J_{\rm eff} \approx 3/2$ ground state. The complete sequence of ground states can be seen in the evolution of $\mu_{\rm eff}(0)$ for $\tilde{U} = 1.4$, shown in Fig.~\ref{fig:d25}(c). For intermediate values of $\tilde{U}$, the $J_{\rm eff} \approx 3/2$ ground state shows a suppressed effective moment, with $\mu_{\rm eff}^{ab} = 0$. This reflects the fact that this doublet is essentially composed of $m_J = \pm 3/2$ states.  These Ising-like moments acquire a finite $\mu_{\rm eff}^c$ due to Hund's coupling, but the transverse moment is essentially zero. We expect that Os-based dimers may lie near the border of $J_{\rm eff} \approx 3/2$ region and $J_{\rm eff} \approx 1/2$ region. 

\subsection{Comparison to Experiment}

\begin{figure}[t]
\includegraphics[angle=0,width=\linewidth]{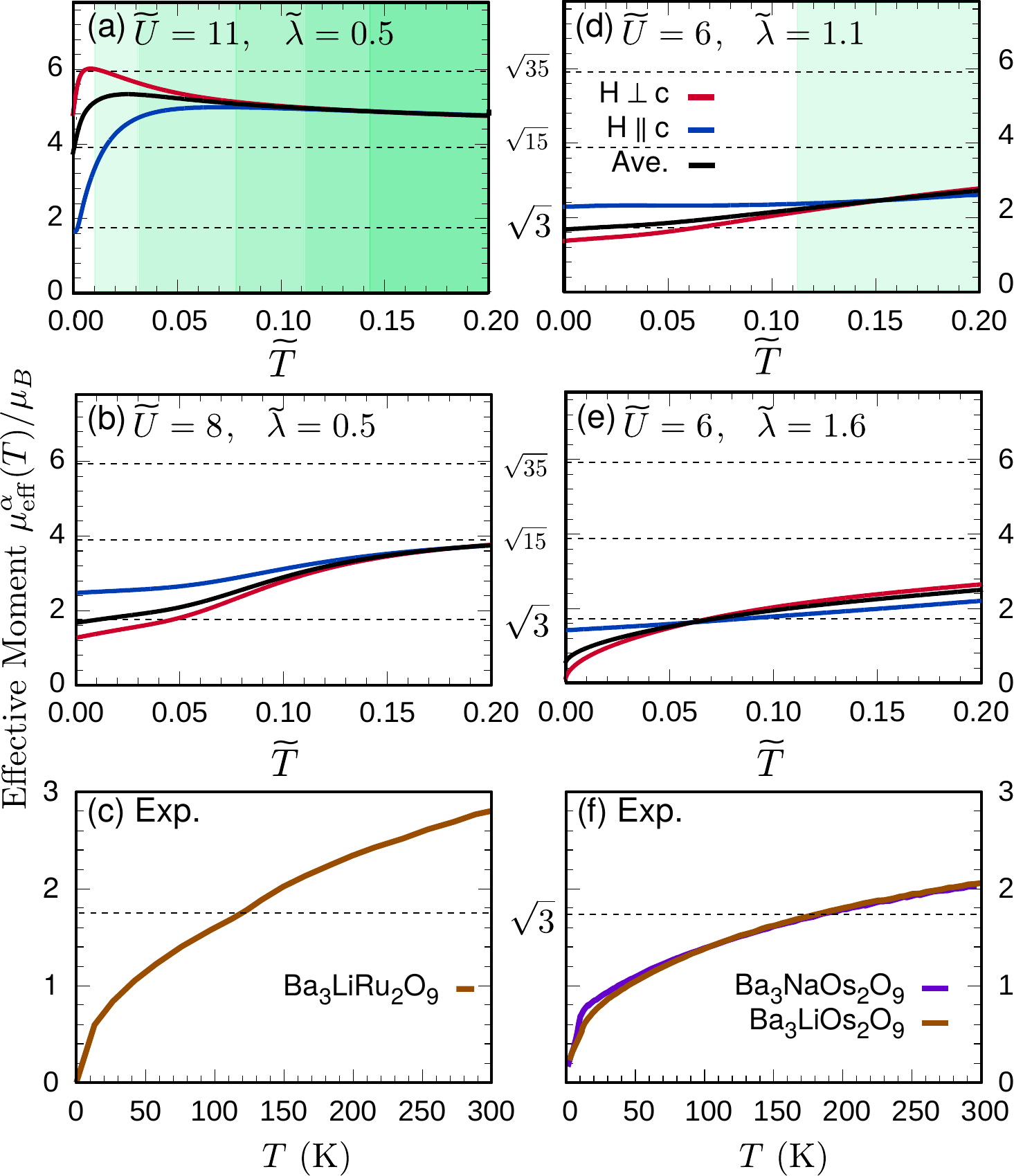}
\caption{Temperature dependence of the magnetic moments for $d^{2.5}$ dimers with various $\tilde{U} (U/t_2)$ and $\tilde{\lambda} ({\lambda/t_2})$ values for $t_1$ = $\Delta$ = 0. (a,b) Theoretical results for isolated $4d$ dimers in high-spin and low-spin ground states, respectively. The green area displays the region
	where excited states are filled.
	(c) Experimental effective moment for Ba$_3$LiRu$_2$O$_9$ based on data from Ref.~\onlinecite{Stitzer2002}. 
(d,e) Theoretical results for isolated $5d$ dimers in $J_{\rm eff} \approx 1/2$ and $J_{\rm eff} \approx 3/2$ ground states, respectively. The energy of the first excited state is above 0.2 $t_2$. (f) Experimental effective moment for Ba$_3$(Na/Li)Os$_2$O$_9$ derived from Ref.~\onlinecite{Stitzer2003}.}
\label{fig:d25T}
\end{figure}

In Fig.~\ref{fig:d25T}, we show the computed temperature dependence of the
effective moment for selected parameters corresponding to regions expected for
real materials. In the strongly relativistic region relevant for Os-based
dimers, the anisotropy of $\mu_{\rm eff}$ is generally weak. In
Fig.~\ref{fig:d25T}(d) and (e), we show the temperature dependence of $\mu_{\rm
eff}$ for $\tilde{\lambda} = 1.1$ and $1.6$, corresponding to the $J_{\rm eff}
\approx 1/2$ and $J_{\rm eff} \approx 3/2$ ground states, respectively. The
green area in Fig.~\ref{fig:d25T}(d)
shows the region where the first excited state is filled.
The energy of the first excited state in Fig.~\ref{fig:d25T}(e)
is above 0.2 $t_2$. For the
latter case, the suppressed $\mu_{\rm eff}^{ab}$ discussed above is observed
only at relatively low temperatures, below room temperature ($\tilde{T} = T/t_2
\sim 0.1$). The theoretical results can be compared with the experimental
$\sqrt{(3k_BT/N_A)\chi_{\rm avg}(T)}$ for Ba$_3$BOs$_2$O$_9$ (B = Na, Li) based
on data from Ref.~\onlinecite{Stitzer2003} shown in Fig.~\ref{fig:d25T} (f). In
principle, the experimental powder data may be consistent with both possible
ground states; single-crystal measurements may be more valuable in the future. 

For Ru-based dimers, there is a stark contrast between the behavior of the high-spin $S \approx 5/2$ (Fig.~\ref{fig:d25T}(a)) and low-spin $J_{\rm eff} \approx 1/2$ (Fig.~\ref{fig:d25T}(b)) cases. The former case shows a non-monotonic average effective moment, with significant anisotropy at low temperatures. At high temperatures, the average value is somewhat suppressed below the spin-only value of $\sqrt{35}\mu_B$ due to SOC. In contrast, in the low-spin case, $\mu_{\rm eff}^\alpha (T)$ evolves monotonically, is more weakly anisotropic, and has average value comparable to a pure $J_{\rm eff} = 1/2$ moment of $\sqrt{3}$. These results can be compared to the experimental $\mu_{\rm eff}(T)$ of Ba$_3$LiRu$_2$O$_9$, shown in Fig.~\ref{fig:d25T}(c), which is based on measurements of the powder susceptibility in Ref.~\onlinecite{Stitzer2002}. For the entire temperature range, the average experimental effective moment remains on the order of $\sqrt{3}$, which suggests that either the material is in a low-spin state, or has a strongly suppressed moment due to strong interdimer couplings or incipient charge order.

\section{Summary and Conclusions} \label{conclusions}
In this work, we considered the theoretical phase diagrams and magnetic response of face-sharing dimers of $4d$ and $5d$ elements with fractional fillings $d^{2.5}$ to $d^{4.5}$. On the basis of these studies we draw various conclusions: 

(i) For these complex materials with competing Coulomb, hopping, and SOC terms,
the magnetic response of individual dimers exhibits strong departures from
conventional Curie behavior. As a result, Curie-Weiss analysis of the magnetic
susceptibility may not be applicable or, if applied, will return strange
results that do not reflect true nature of the local magnetic moment. Except
for the most standard case of large $\tilde U$ and small $\tilde\lambda$,
essentially all interaction regimes considered in our study lead to effective
moments that are strongly temperature-dependent up to at least $\tilde T=0.1$
that corresponds to room temperature and represents the typical range of the
Curie-Weiss fitting of the experimental data. In this situation, analyzing
temperature-dependent effective moment can be a better strategy that yields
valuable information regarding the local electronic ground states of the
dimers. The data above room temperature and experimental information on the
anisotropy will strongly improve the accuracy of such an analysis.

(ii) The ground states of the dimers depend on the relative values of Coulomb
interactions $J_H, U$, spin-orbit coupling strength $\lambda$, and intradimer
hoppings $t_1$, and $t_2$. An important consideration is that increasing
$\lambda$ leads to several crossings of single-particle levels of different
symmetry, which manifests in phase transitions between different ground-state
configurations depending on the specific filling. From the experimental
perspective, this level crossing leads to electronic excitations lying within
the energy range of $50-100$\,meV and accessible for inelastic neutron
scattering, which can provide additional valuable information on the electronic
structure. It also justifies the application of empirical models that were used
to describe magnetic susceptibility of mixed-valence dimers in terms of
excitations between several electronic states~\cite{Ziat2017,Dey2017}. A caveat
here is that none of these states feature purely spin moments, so their
energies and magnetic moments should be both treated as fitting parameters,
with the risk of making the model overparameterized.

(iii) For all considered fillings, $4d$ Ru- and Rh-based dimers may, in principle, exhibit either low-spin or high-spin ground states. From a survey of available experimental susceptibility data, we find that the vast majority of such materials likely fall into a low-spin ground state. Provided $C_3$ symmetry of the dimers is maintained, all low-spin states have unquenched orbital moments, which strongly modify the magnetic response via coupling to the $S=1/2$ degrees of freedom. Thus, consideration of SOC is essential for these materials. Intradimer charge order, previously suggested to exist in some of these materials, is unnecessary to explain the strong suppression of the effective magnetic moments. 

(iv) For $5d$ materials, various ground states are possible, having either pure $J_{\rm eff} = 3/2$ character, or mixed $J_{\rm eff} = 1/2$ and $J_{\rm eff} = 3/2$ character. Identifying the nature of these local degrees of freedom is an essential prerequisite for establishing minimal magnetic models for these materials. For the case of the spin-liquid candidate material Ba$_3$InIr$_2$O$_9$ with $d^{4.5}$ filling of the Ir sites, the magnetic response is suggestive of a ``low-spin'' $J_{\rm eff} \approx 3/2$ ground state. Once the ground state configuration can be further confirmed, for example via RIXS, analysis of the interdimer couplings would be of significant interest. The temperature-dependence of the susceptibility can be modelled as arising from predominantly single-dimer effects, which would be consistent with relatively weak interdimer magnetic couplings. 

We hope that this discussion provides a comprehensive overview of the properties of face-sharing $4d$ and $5d$ dimers with fractional fillings, which may serve as a starting point for future experimental and theoretical studies of these complex materials. 

\acknowledgments
 R.V., S.W. and Y.L. thank Daniel
Khomskii, Igor Mazin, Sergey V. Streltsov, Markus Gr\"uninger and Adam A. Aczel for discussions and acknowledge the support by the Deutsche
Forschungsgemeinschaft (DFG) through grant VA117/15-1.
Computer time was allotted at the
Centre for for Scientific Computing (CSC) in Frankfurt.
R.V. and S.W.  acknowledge support from
the National Science Foundation under Grant No.
NSF PHY-1748958 and hospitality from KITP where part of this work
was performed.  Y.L. acknowledge support from the Fundamental Research Funds for the Central Universities (Grant No. xxj032019006)
and China Postdoctoral Science Foundation (Grant No. 2019M660249). The work in Augsburg
was supported by DFG under TRR80 and by the Federal Ministry for Education and
Research through the Sofja Kovalevskaya Award of Alexander von Humboldt
Foundation (A.A.T.)

$*$ yingli1227@xjtu.edu.cn\\
$\dagger$ Current address: Department of Physics, Indian Institute of Technology (Indian School of Mines), Dhanbad, Jharkhand, 826004, India \\ 
$\ddagger$ valenti@itp.uni-frankfurt.de \\
$\mathsection$ winter@physik.uni-frankfurt.de

\bibliography{ref}

\appendix

\section{Character Table for $D_{3h}$ Double Group} \label{app:d3hdouble}

\begin{table}[ht]
\caption {Partial character table for $D_{3h}$ Double Group}
\label{tab:d3hdouble}
\begin{ruledtabular}
\begin{tabular}{r@{\hspace{2em}}rrrrrrrrr}
& $E$& $C_3 $ & $3 C_2^{\prime}$ & ${\sigma}_h$ & $S_6$ & $3{\sigma}_v$ & $\overline{E}$ & $\overline{C}_3$ & $\overline{S}_6$\\
\hline
$E_{1/2}$& 2 & 1 & 0 & 0 & $\sqrt{3}$ & 0 & $-2$ & $-1$ & $-\sqrt{3}$ \\
$E_{3/2}$& 2 & $-2$ & 0 & 0 & 0 & 0 & $-2$ & 2 & 0 \\
$E_{5/2}$& 2 & 1 & 0 & 0 & $-\sqrt{3}$ & 0 & $-2$ & $-1$ & $\sqrt{3}$ \\
\end{tabular}
\end{ruledtabular}
\end{table}

\section{Details of crystal growth and characterization for Ba$_3$InIr$_2$O$_9$} \label{app:prep}

Small single crystals of {\bainiro} were grown from the pre-reacted
polycrystalline material in the BaCl$_2$ flux taken in the 1:10 ratio. The
mixture was heated to 1200$^{\circ}$\,C in 9\,hours and kept at this
temperature for 20\,hours followed by a slow cooling to 950\,$^{\circ}$ in 90
hours and a standard furnace cooling to room temperature. The resulting
crystals had platelet hexagonal form and linear dimensions of less than
0.3\,mm. They showed same lattice parameters and hexagonal symmetry as the
powder samples reported previously~\cite{Dey2017}. Energy-dispersive x-ray
spectroscopy (EDXS) analysis revealed the element ratio of In:Ir:Ba =
1:2.08(6):3.17(9) in good agreement with the Ba$_3$InIr$_2$O$_9$ composition.
The $c$-direction of the crystals was determined by monitoring reflections on
the x-ray powder diffractometer (Rigaku MiniFlex, CuK$_{\alpha}$ radiation).

\begin{figure}
\includegraphics{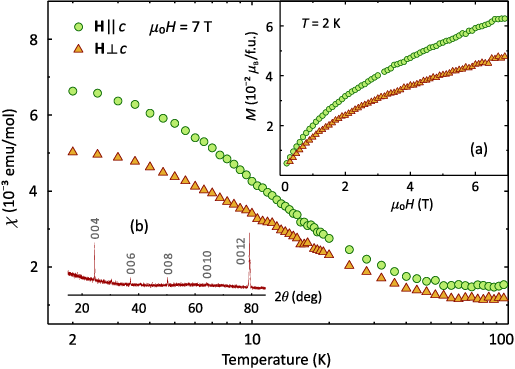}
\caption{\label{fig:exp-chi}
Temperature-dependent magnetic susceptibility measured on the Ba$_3$InIr$_2$O$_9$ single crystal in the applied field of 7\,T for different field directions. Inset (a): field-dependent magnetization at 2\,K. Inset (b): x-ray diffraction pattern from the $[001]$ surface of the crystal.
}
\end{figure}

Magnetic susceptibility (Fig.~\ref{fig:exp-chi}) was measured on a single
crystal using the MPMS III SQUID from Quantum Design in the temperature range
of $1.8-200$\,K in the applied field of 7\,T. Above 200\,K, the signal was
below the sensitivity limit, owing to the very small sample size. As the sample
mass could not be determined with sufficient accuracy, the data measured for
$H\,\|\,c$ and $H\!\perp\!c$ were averaged and scaled against the powder data.
The powder data were measured up to 650\,K, as explained in
Ref.~\onlinecite{Dey2017}.

\end{document}